\documentclass[pra,aps,amsfonts,amsmath,superscriptaddress,showpacs,twocolumn]{revtex4}
\usepackage{graphicx}

\begin{document}
\title{An environment-mediated quantum deleter}
\author{R. Srikanth}
\email{srik@rri.res.in}
\affiliation{Poornaprajna Institute of Scientific Research, 
Devanahalli, Bangalore-562 110, India.} 
\affiliation{Raman Research Institute, Sadashiva Nagar, Bangalore- 560 080,
India.}
\author{Subhashish Banerjee}
\email{subhashishb@rri.res.in}
\affiliation{Raman Research Institute, Sadashiva Nagar, Bangalore- 560 080,
India.}
\begin{abstract}
Environment-induced decoherence presents a great challenge
to realizing a quantum computer. We point out the somewhat surprising
fact that decoherence can be useful, indeed necessary, for practical
quantum computation, in particular, for the effective erasure of quantum
memory in order to initialize the state
of the quantum computer. The essential point behind the deleter
is that the environment, by means of a dissipative interaction, furnishes a 
contractive map towards a pure state.
We present a specific example of an amplitude damping channel provided by
a two-level system's interaction with its environment
in the weak Born-Markov approximation. This is contrasted with a purely
dephasing, non-dissipative channel provided by a two-level system's
interaction with its environment by means of 
a quantum nondemolition interaction.
We point out that currently used state preparation techniques, for example 
using optical pumping, essentially perform as quantum deleters.
\end{abstract}

\maketitle 

Quantum computation is  well known to solve certain  types of problems
more  efficiently  than classical  computation  \cite{nc00}. A
basic challenge facing the realization of a
quantum   computer  is  that   of  fighting
decoherence. Although quantum mechanical linearity endows a quantum computer
with greater-than-classical power \cite{shor}, 
it  also imposes certain restrictions, such as
the prohibition on perfect cloning  \cite{woo82} and on deleting
the copy of an arbitrary quantum state perfectly \cite{nodel}. 

A quantum computational task can be broadly divided into three stages:
(1) initializing the quantum computer, by preparing all qubits in
a standard `blank state'; (2) executing the unitary operation that
performs the actual computation; (3) performing measurements to read off
results. 

The well known difficulty in realizing a quantum computer is that
of shielding it from the environment during step (2) 
\cite{cot06}. This is the
problem of fighting decoherence, the loss of coherence in a system
(here a qubit), due to interaction with
its environment, which has been the
subject of intense research. A variety of techniques, including quantum
error avoidance \cite{zan97}, quantum
error correction \cite{css}, quantum error prevention
\cite{erprev}, dynamic decoupling \cite{dyndec}, frequency modulation
of the heat bath \cite{gsa00},
fault tolerant quantum computation \cite{ftqc}, 
decoherence-free subspaces \cite{dfss}, among others, exist 
to combat decoherence.
Our main aim here, however, is to point out that in one of the stages of
quatum computation, decoherence is useful, even necessary.

In step (1), we must be able to erase quantum memory 
of the state inherited from a previous computational task,
in order to prepare the state of a quantum computer for a subsequent
task.  What is  required is a  quantum mechanism that
with high probability  allows us to prepare standard  `blank states',
usually designated by the pure state $|0\rangle$. 
It is clear that no unitary process can achieve this, since true deletion
would be  irreversible, and hence  non-unitary. Given two distinct
states $|\psi_1\rangle$ and $|\psi_2\rangle$, a purported 
deleting operation
$\delta$ should effect $\delta |\psi_1\rangle \longrightarrow
|0\rangle$ and $\delta |\psi_2\rangle \longrightarrow
|0\rangle$ \cite{cai04}. Unitarity requires that $\langle\psi_1|\psi_2\rangle =
\langle 0|0\rangle=1$, which cannot be satisfied unless
$\psi_1 = \psi_2$. Further, 
the no-deleting theorem implies that no qubit state can be erased
against a copy \cite{nodel}.

A direct method for initializing the quantum computer would be
to measure all qubits in the computational basis. This results in a
statistical mixture of $|0\rangle$'s and $|1\rangle$'s, but there is no
unitary way in a closed system to flip the $|1\rangle$'s while retaining 
the $|0\rangle$'s. However, open quantum systems 
can effect non-unitary evolution on  a sub-system of
interest. We are thus led to conclude that decoherence is in
fact {\em necessary} for step 1. In particular, decoherence
must furnish a contractive map that drives any initial state of the system
towards a fixed pure state to serve as the blank state.

In this  work, we use  this insight to  argue that decoherence  can be
useful to quantum computation. In  particular,  we  show  that a
dissipative interaction with an environment can  serve  as  an
effective deleter  of quantum information.  Our result is comparable
to that presented in Ref. \cite{rom06}, where the evolution of a
qubit-probe system is modified by interaction with an environment, which
enables the qubit to be driven to a target state, though in function,
their work aims for quantum control, whereas the present proposal is
aimed at state preparation. 

We briefly introduce the interaction of a two-level system (a qubit)
with an environment (bath) of harmonic
oscillators via a dissipative as well as non-dissipative interaction. 
Our model assumes that the qubits of the computer
are mutually non-interacting and decohere independently. 
However, it may be noted that in many situations this is a reasonable
assumption. In fact,
conventional quantum error-correction codes \cite{css} were
developed primarily to deal with independent, incoherent errors.
In the dissipative case the system-environment ($S$-$R$)
interaction is treated in a standard 
Born-Markov approximation. In the non-dissipative
case, the $S$-$R$ interaction is of a quantum nondemolition (QND) type. 
These environmental interactions are representative
of open system effects, and further, correspond to two 
of the most important noisy channels in quantum information
theory. In particular, 
the dissipative type of system-environment interaction
yields the (generalized) amplitude damping channel, while 
the non-dissipative interaction yields the phase damping channel
\cite{nc00}. We point out that whereas the former
is  useful to engineer a quantum  deleter, 
the latter leads to a mixed state, which is unsuitable for state preparation.
This observation is compatible with the fact that erasing information
is an irreversible process that dissipates energy \cite{lan82,pie00},
affirming the connection between thermodynamics and information theory
\cite{ben82}.

The total Hamiltonian is 
$H = H_S + H_R + H_{SR}$,
where $H_S$, $H_R$ and $H_{SR}$ stand for the Hamiltonians of
the system, reservoir and $S$-$R$ interaction,
respectively. Here the system Hamiltonian is given by
$H_S = (\hbar \omega/2) \sigma_3$,
with $\sigma_3$ being the usual Pauli matrix. For the reservoir
Hamiltonian we use the standard form of a bath of harmonic
oscillators, i.e., $H_R = \hbar\omega_k 
b^{\dagger}_k b_k$. We assume separable initial
conditions, i.e.,
$\rho (0) = \rho^s (0) \rho_R (0)$,
and the reservoir is assumed to be initially in a squeezed thermal state, i.e.,
a squeezed thermal bath, with an initial density matrix $\rho_R 
(0)$ given by
\begin{equation}
{\rho}_R(0) = {S} (r,\Phi) {\rho}_{th} 
{S}^{\dagger} (r,\Phi),
\label{eq:reservoir}
\end{equation}
where
\begin{equation}
{\rho}_{th} = \prod_k \left[ 1 - e^{-\beta \hbar \omega_k} 
\right] \exp\left(-\beta \hbar\omega_k b^{\dag}_kb_k\right)
\label{2e} 
\end{equation}
is the density matrix of the thermal bath, and
\begin{equation}
{S} (r_k, \Phi_k) = \exp \left[ r_k \left( {{b}^2_k 
\over 2} e^{-i2\Phi_k} - {{b}^{\dagger 2}_k \over 2} 
e^{i2\Phi_k} \right) \right] \label{2f} 
\end{equation}
is the squeezing operator with $r_k$, $\Phi_k$ being the
squeezing parameters \cite{cs85}. Squeezing of the bath has been
shown to be useful in the suppression of decay of quantum coherence
\cite{kb93}, and to modify the evolution of the geometric phase of 
two-level quantum systems \cite{sribann}.
Hence it is of relevance to study its possible
influence on the behavior of the quantum deleter.

The system-environment interaction is taken to be dissipative
and of the weak Born-Markov type \cite{bp02} leading to a standard
Lindblad equation, which in the interaction
picture has the following form \cite{sribann} 
\begin{equation}
\frac{d}{dt}\rho^s(t) = \sum_{j=1}^2\left(
2R_j\rho^s R^{\dag}_j - R_j^{\dag}R_j\rho^s - \rho^s R_j^{\dag}R_j\right),
\end{equation}
where $R_1 = (\gamma_0(N_{\rm th}+1)/2)^{1/2}R$,
$R_2 = (\gamma_0N_{\rm th}/2)^{1/2}R^{\dag}$ and 
$N_{\rm th} = (\exp(\hbar\omega/k_B T) - 1)^{-1}$,
is the Planck distribution giving the number of thermal
photons at the frequency $\omega$. Here 
$R = \sigma_-\cosh(r) + e^{i\Phi}\sigma_+\sinh(r)$, and the quantities
$r$ and $\Phi$ are the environmental squeezing parameters and
$\sigma_{\pm} = \frac{1}{2}\left(\sigma_1 \pm i\sigma_2\right)$.
If $T=0$, so that $N_{\rm th}=0$, then $R_2$ vanishes, and a single
Lindblad operator suffices.

From this, the Bloch vectors can be obtained as \cite{sribann}
\begin{eqnarray}
\label{eq:2b0}
\langle \sigma_1 (t) \rangle &=& \left[1 + \frac{1}{2} \left(e^{\gamma_0 a t}
- 1\right) (1 + \cos(\Phi))\right] \beta(t)
\langle \sigma_1 (0) \rangle \nonumber\\
&-& \sin(\Phi) \sinh({\gamma_0 a t \over 2}) e^{-{\gamma_0 \over 2}(2N + 1)t}
\langle \sigma_2 (0) \rangle, \nonumber \\
\langle \sigma_2 (t) \rangle &=& \left[1 + {1 \over 2} \left(e^{\gamma_0 a t}
- 1\right) (1 - \cos(\Phi))\right] \beta(t)
\langle \sigma_2 (0) \rangle \nonumber\\
&-& \sin(\Phi) \sinh({\gamma_0 a t \over 2}) e^{-{\gamma_0 \over 2}(2N + 1)t}
\langle \sigma_1 (0) \rangle, \nonumber \\
\langle \sigma_3 (t) \rangle &=& e^{-\gamma_0 (2N + 1)t} \langle 
\sigma_3 (0) \rangle - \frac{\left(1 - e^{-\gamma_0 (2N + 1)t} 
\right)}{2N+1}, 
\end{eqnarray}
where $\beta(t) \equiv \exp(-{\gamma_0 \over 2}(2N + 1 + a)t)$,
$a = \sinh(2r) (2N_{\rm th} + 1)$,
$N = N_{\rm th}(\cosh^2(r) + \sinh^2(r)) + \sinh^2(r)$, 
and $\gamma_0$ is a constant typically denoting the system-environment 
coupling strength, while $\omega$ is the system frequency.
In Eqs. (\ref{eq:2b0}), the quantities $\langle \sigma_j(0)\rangle$ 
$(j=1,2,3)$ are the expectation values of the respective Pauli
operators with respect to the intial state
\begin{equation}
|\psi(0)\rangle = \cos\left(\frac{\theta_0}{2}\right) |1\rangle +
e^{i \phi_0}
\sin\left(\frac{\theta_0}{2}\right) |0\rangle.
\label{eq:ini}
\end{equation} 

It is seen from Eqs. (\ref{eq:2b0}) that for  fixed
squeezing, the asymptotic equilibrium state $\rho_{\rm asymp}$,
given by
\begin{equation}
\label{eq:asymp}
\rho_{\rm asymp} \equiv \left(\begin{array}{ll} 1-p & 0 \\
0 & p
\end{array} \right),
\end{equation}  
where $p = \frac{1}{2}\left[1 + \frac{1}{2N+1}\right]$,
is approached faster for stronger coupling, as might be expected. The
entire Bloch  sphere shrinks towards $\rho_{\rm asymp}$.
In the absence of squeezing and at zero temperature ($T$), this action 
corresponds exactly to an 
amplitude damping channel, for which $N=0$ and thus 
$p=1$. This corresponds to the
point representing the state $|0\rangle$ (the Bloch sphere south pole 
in our notation).  
For the case of finite $T$ but zero squeezing, this corresponds
to a generalized
amplitude damping channel, for which $N > 0$, giving
$p < 1$, i.e., a mixed state \cite{sribann}. 
In the limit of infinite temperature, $N \mapsto \infty$ and
$p \mapsto \frac{1}{2}$, so that $\rho_{\rm asymp}$ tends to the maximally
mixed state, represented by the center of the Bloch sphere.
Thus, the interaction with the environment provides a contractive map,
such that the asymptotic state is pure ($p=1$) or mixed ($p < 1$), 
depending on environmental conditions.

It is worth stressing that the quantum deleter requires a
dissipative interaction with its environment. 
This may be contrasted with a nondissipative interaction, in order 
to shed light on why the former is necessary for our purpose. 
In the case of a non-dissipative, quantum nondemolition (QND)
interaction, the environment acts as
a purely quantum dephasing channel that leaves the 
energy of the system unchanged \cite{sribann}. This is illustrated
briefly in Appendix \ref{sec:p}.
\begin{figure}
\includegraphics[width=7cm]{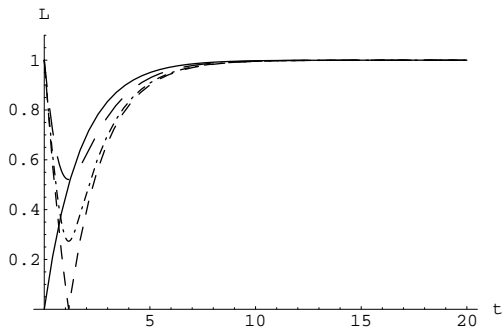}
\caption{Evolution of the length $L$ of the Bloch vector (Eqs. (\ref{eq:2b0}))
for various initial states, with $\gamma_0 = 0.5$, $\omega=1.0$,
temperature  $T=0$ and the squeezing parameters set to  zero.
The small-dashed, dot-dashed and large-dashed curves represent
initial pure states with $\theta_0 = 0$, $\pi/8$ and $\pi/4$,
respectively (in this case, there is no dependence of $L$ on azimuthal
angle $\phi_0$). The solid curve
represents the intially maximally mixed state.
In the case of the pure states, note that after at first
becoming mixed  on account of  entanglement with the  environment, the
qubit returns to purity (in the state $|0\rangle$).}
\label{fig:rbma}
\end{figure}

Fig. \ref{fig:rbma} depicts the length of the Bloch vector 
from Eqs. (\ref{eq:2b0})
for various initial states (both pure as well as maximally mixed)
as a function of time. It can be shown 
from Eqs. (\ref{eq:2b0}) that for all pure states 
except $|0\rangle$, the quantity ${\bf r}\cdot d{\bf r}/dt$,
where ${\bf r}$ is the Bloch vector, is negative at $t=0$, implying that
the tip of the Bloch vector plunges initially into the Bloch sphere 
on its way towards the point representing the stationary state $|0\rangle$.
This is reflected in the initial reduction of
the length of the Bloch vector, during which time, the  qubit   becomes
increasingly entangled  with the environment, before gradually
factoring out.

The proposed quantum  deleter works as follows.  To clear  the memory of the
quantum computer, the control processes brought into play to shield
it from environmental decoherence, are 
turned off for a short time
$t$, during which each qubit in the quantum computer 
is assumed to
interact independently with and equilibriate with its environment,
assumed to be an unsqueezed vacuum bath.
We may characterize the performance of the quantum deleter in
terms  of fidelity,  a  measure  of closeness  of  two quantum  states
\cite{nc00}.   From  Eqs. (\ref{eq:2b0}),   we  find   that  with   probability
exponentially approaching  unity, the  fidelity 
$f(t) \equiv F(\rho^s(t),|0\rangle)$ of a  qubit approaches
the designated standard `blank state' $|0\rangle$ according to
\begin{eqnarray}
\label{eq:fid}
f(t) &=& \sqrt{\langle 0|\rho^s(t)|0\rangle} = 
\sqrt{{1 - \langle \sigma_3 (t) \rangle \over 2}} \nonumber\\
&=& \frac{1}{\sqrt{2}}\left[\left(1 - e^{-\Gamma t}
\langle \sigma_3(0)\rangle
\right)
+ \frac{\left(1 - e^{-\Gamma t}\right)}{2N+1}\right]^{1/2}
\end{eqnarray}
where $\Gamma \equiv \gamma_0 (2N + 1) $ and
$\langle  \sigma_3(0)\rangle$  is  the expectation  value  of
$\sigma_3$ at  time $t=0$.  

Observe that at any time $t$, fidelity is smallest when 
$\langle \sigma_3(0)\rangle$ is largest, i.e., 1. Thus, no matter
what the initial state, we can lower bound fidelity as
\begin{equation}
\label{eq:lowbou}
1 \ge f(t) \ge 
\frac{1}{\sqrt{2}}\left[\left(1 - e^{-\Gamma t}\right)\left(1 +
\frac{1}{2N+1}\right)\right]^{1/2}.
\end{equation}
Observe that if and only if temperature and squeezing are zero, 
implying $N=0$, the asymptotic value of $f(t)$ is 1, i.e., $f(\infty)
= 1$. If any of these two conditions is not met, then $N > 0$ and hence
$f(\infty) < 1$. Therefore, given zero temperature and zero squeezing,
one may select the `off-shield' duration $t$ that 
guarantees fidelity equal to or higher than any pre-selected value
(less than 1).

\begin{figure}
\includegraphics[width=7.0cm]{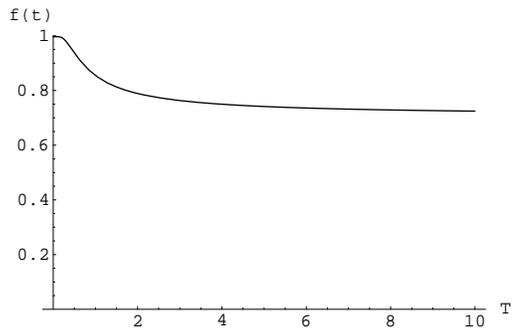}
\caption{Fidelity ($f(t)$) falls as a function of temperature ($T$,
in units where $\hbar \equiv k_B \equiv 1$) until it
reaches the value $1/\sqrt{2}$ corresponding to a maximally mixed state.
The case shown here corresponds to $\theta_0 = 0$, 
$\gamma_0 = 0.5$, $\omega=1.0$ and time $t=10$. Here we set the squeezing
parameters $r$ and $\Phi$ to zero.}
\label{fig:fidtemp}
\end{figure}

\begin{figure}
\includegraphics[width=7cm]{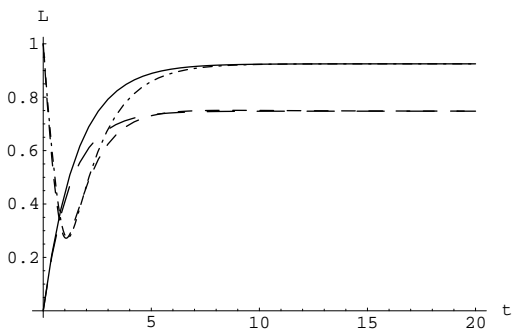}
\caption{The effect of squeezing: The bold and the dot-dashed lines
correspond to $r=0.2$, the latter representing an initial pure
state with $\theta_0=\pi/4$, the former an initially maximally mixed state.
The large-dashed (maximally mixed initially) 
and small-dashed (initially pure state with
$\theta_0=\pi/4$) lines represent the evolution of the Bloch vector
for the same initial states, but with $r=-0.4$.
In all cases, $T=0$, $\omega=1.0$ and $\gamma_0=0.6$.} 
\label{fig:rbma0}
\end{figure}

As seen from  Eq.  (\ref{eq:asymp}),  the  effect of increasing
temperature is  to 
make  the   asymptotic  state $\rho_{\rm asymp}$ ever more
mixed.  As  $T   \rightarrow  \infty$,
$\rho_{\rm asymp}$  tends to  the  maximally mixed  state.
Fig. \ref{fig:fidtemp} depicts
the effect of increasing temperature on fidelity.
Fig. \ref{fig:rbma0} depict the effect of squeezing on the evolution
of the Bloch vector length (from Eqs. (\ref{eq:2b0})). It can be seen that
due to the presense of squeezing, the system does not return to 
purity in the state $|0\rangle$. Therefore,
for optimal performance of the quantum deleter, i.e., for 
high purity of the output state of the system qubit,
both the squeezing parameter and temperature  must be set 
as close to zero as possible (i.e.,
$T=0$ and $r=0$, yielding $N = 0$ and hence $p=1$). 

It is interesting to note that a contractive map finds another
practical, but quite different (cryptographic), use.
In the universal quantum homogenizer
(UQH), proposed by Ziman {\em et al.} \cite{zim02}, a 
sequence of interactions
of a system qubit with a reservoir of qubits, initially prepared
in an identical state $\xi$,
drives an arbitrary state towards $\xi$. As
a result, the UQH acts as a {\em quantum safe} with a classical key,
consisting of the interaction sequence. One might also consider the
UQH being turned into a quantum deleter by setting $\xi \equiv |0\rangle$.
However this would require a sequence of highly controlled operations,
which would render it relatively difficult to implement, in comparison 
to allowing a system to cool via interaction with its environment, as
required in our case.

We believe that the most interesting aspect of our work is the idea
that the unavoidable problem posed by decoherence is shown to be
not only useful, but indeed necessary for quantum computation, in
particular for initial state preparation. By allowing a role to be
played by a realistic environment, in particular, one in which
squeezing is absent, it improves the chance that a
quantum computer can be practically realized. In practice, 
what is needed is to turn off in step (1) control processes used
to protect the quantum information in step (2) against decoherence 
due to a vacuum environment.  For example, in a quantum information
processing system using ultracold atoms in a magneto-optic trap,
one may allow spontaneous decay of excitations. However, 
one should not switch off the trap, since this would 
effectively bring the system in contact with a finite temperature 
environment.

It would at first appear that the requirement of a vacuum bath ($T=0$) 
places a technological hurdle. However, we note that state
preparation techniques in atom-optical experiments, quantum dots, etc.
essentially involve a quantum deletion process, 
usually furnished by a spontaneous
emission. Some examples are, preparation of a Zeeman-state in
an optical lattice by two-dimensional sideband Raman cooling
\cite{tai01}, and by laser cooling of the spin of an electron 
trapped in a semiconductor quantum dot \cite{ata06}.
In this context, the notable experiment by Myatt {\em et al.} 
\cite{mya00}, in which they
engineered a (nearly) zero-temperature reservoir, is worth pointing out.


We are thankful to Prof. Hema Ramachandran for useful discussions.

\appendix
\section{Evolution governed by QND $S$-$R$ interaction\label{sec:p}}
We consider a system interacting with its bath via
a QND interaction. The Hamiltonian is 
\begin{equation}
H =  H_S + \sum\limits_k \hbar \omega_k b^{\dagger}_k b_k + H_S 
\sum\limits_k g_k (b_k+b^{\dagger}_k) + H^2_S \sum\limits_k 
{g^2_k \over \hbar \omega_k}. \label{2a} 
\end{equation} 
The second term on the RHS of the above equation
is the free Hamiltonian of the environment, while the third
term is the $S$-$R$ interaction Hamiltonian.
The last term on the RHS of Eq. (\ref{2a})
is a renormalization inducing `counter term'. Since $[H_S, 
H_{SR}]=0$, Eq. (\ref{2a}) is of QND type.

Following Ref. \cite{sribann},
taking into account the effect of the
environment modelled as a squeezed thermal bath
(Eqs. (\ref{eq:reservoir}), (\ref{2e}) and (\ref{2f})),
the reduced dynamics of the system can be shown to be
\begin{widetext}
\begin{equation}
\rho^s_{nm} (t)  =  e^{-{i \over \hbar}(E_n-E_m)t} e^{
i(E^2_n-E^2_m) \eta(t)}
e^{- (E_n-E_m)^2 \gamma(t)} 
\rho^s_{nm} (0), \label{2g} 
\end{equation}
\end{widetext}
where the explicit forms of $\eta(t)$ and $\gamma(t)$ can be
obtained from Ref. \cite{sribann}. Here $E_n$ is the eigenstate
of the system Hamiltonian defined in the system eigenbasis.
It can be seen from the above equation that  whereas the off-diagonal
elements of the reduced density matrix decay with time, the diagonal
elements remain unaffected. Clearly, this feature makes a QND $S$-$R$ 
interaction unsuitable for quantum deletion. We note that 
Eq. (\ref{2g}) does not depend on the specific form of the system
Hamiltonian. 

For the case of a two-level system, Eq. (\ref{2g}) can be recast
in the form of Bloch vectors as follows \cite{sribann}:
\begin{eqnarray}
\label{eq:2g0}
\langle \sigma_1 (t) \rangle &=& \sin(\theta_0) \cos(\omega t + \phi_0)
e^{-(\hbar \omega)^2 \gamma(t)}, \nonumber \\
\langle \sigma_2 (t) \rangle &=& \sin(\theta_0) \sin(\omega t + \phi_0)
e^{-(\hbar \omega)^2 \gamma(t)}, \nonumber \\
\langle \sigma_3 (t) \rangle &=& \cos(\theta_0), \label{2i}  
\end{eqnarray}
where $\gamma = \gamma(\gamma_0,T,r,\Phi) \ge 0$ ($\gamma=0$ if and only
if the environmental interaction is absent) 
is calculated for the $T=0$ and high $T$ case in Ref. \cite{sribann}.
It suffices for our present purpose to note that the action described by
Eqs. (\ref{eq:2g0}) is that of contracting the Bloch sphere along 
the $\sigma_3$-axis. An initial pure state is driven to
a mixed state that is diagonal in the computational basis, {\em even
at $T=0$}.  Thus, this is not useful for our purpose.


\begin{thebibliography}{100}
\bibitem{nc00} M. Nielsen and I. Chuang, {\em Quantum Computation and
Quantum Information}, Cambridge (2000).
\bibitem{shor} P. W. Shor, SIAM J. Sci. Statist. Comput. {\bf 26}, 1484 (1997);
eprint quant-ph/9508027; L. K. Grover, Phys. Rev. Lett. {\bf 79}, 325 (1997).
\bibitem{woo82} W. K. Wooters and W. H. Zurek, Nature {\bf 299}, 802 (1982).
\bibitem{nodel} A. K. Pati and S. Braunstein, Nature {\bf 404}, 164 (2000).
\bibitem{cot06} R. C\^ot\'e, Nature Physics {\bf 2}, 583 (2006);
A. Andr\'e, D. Demille, J. M. Doyle, {\em et al.}, {\em ibid.,}
636 (2006).
\bibitem{zan97} P. Zanardi and 
M. Rasetti, Phys. Rev. Lett. {\bf 79}, 3306 (1997).
\bibitem{css} P. W. Shor, Phys. Rev. A {\bf 52}, R2493 (1995);
A. R. Calderbank and P. W. Shor, Phys. Rev. A {\bf 54}, 1098 (1996);
A. Steane, Proc. Roy. Soc., London, Ser. A {\bf 452}, 2551 (1996).
\bibitem{erprev} L. Vaidman, L. Goldenberg, and S. Wiesner
Phys. Rev. A {\bf 54}, R1745 (1996).
\bibitem{dyndec} L. Viola and S. Lloyd, Phys. Rev. A {\bf 58}, 2733 (1998);
D. Vitali and P. Tombesi, Phys. Rev. A {\bf 65}, 012305 (2001).
\bibitem{gsa00} G. S. Agarwal, Phys. Rev. A {\bf 61}, 013809 (1999);
G. S. Agarwal, M. O. Scully and H. Walther, Phys. Rev. Lett. {\bf 86},
4271 (2001).
\bibitem{ftqc} D. Gottesman, Phys. Rev. A {\bf 57}, 127 (1998).
\bibitem{dfss} D. A. Lidar, I. L. Chuang and K. B. Whaley,
Phys. Rev. Lett. {\bf 81}, 2594 (1998).
\bibitem{cai04} Q.-Y. Cai, Chin. Phys. Lett. {\bf 21}, 1189 (2004);
eprint quant-ph/0401005.
\bibitem{rom06} R. Romano and D. D'Alessandro, Phys. Rev. Lett. {\bf 97},
080402 (2006).
\bibitem{lan82} R. Landauer, IBM J Res. Dev. {\bf 5}, 183 (1961).
\bibitem{pie00} B. Piechocinska, Phys. Rev. A {\bf 61}, 062314 (2000).
\bibitem{ben82} C. H. Bennett, Int. J. Theor. Phys. {\bf 21}, 905 (1982).
\bibitem{cs85} C. M. Caves and B. L. Schumaker, Phys. Rev. A 
{\bf 31}, 3068 (1985); B. L. Schumaker and C. M. Caves, Phys. 
Rev. A {\bf 31}, 3093 (1985). 
\bibitem{kb93} M. S. Kim and V. Bu{\v z}ek, Phys. Rev. A {\bf 47},
610 (1993).
\bibitem{bp02} H. -P. Breuer and F. Petruccione, 
{\it The Theory of Open Quantum Systems} (Oxford University Press, 2002).
\bibitem{sribann} S. Banerjee and R. Srikanth, eprint quant-ph/0611161.
\bibitem{zim02} M. Ziman, P. \v{S}telmachovi\v{c}, V. Bu\v{z}ek, M. Hillery,
V. Scarani and N. Gisin, Phys. Rev. A {\bf 65}, 042105 (2002).
\bibitem{tai01} A. V. Taichenachev, A. M. Tumaikin, and V. I. Yudin
and L. Hollberg, Phys. Rev. A {\bf 63}, 033402 (2001).
\bibitem{ata06} M. Atat\"ure, J. Dreiser, A. Badolato, {\em et al.},
Science {\bf 312}, 551 (2006).
\bibitem{mya00} C. J. Myatt, B. E. King, Q. A. Turchette, {\em et al.},
Nature {\bf 403}, 269 (2000).
\end{thebibliography}
\end{document}